\documentclass[pra]{revtex4}
 \usepackage{amssymb} \usepackage{graphicx}

\begin{document}
 \title{Higher-degree Dirac Currents of Twistor and Killing Spinors in Supergravity Theories}
\author{\"{O}zg\"{u}r A\c{c}{\i}k}
\email{ozacik@science.ankara.edu.tr}
 \author{\"Umit Ertem}
 \email{umitertemm@gmail.com}
\address{Department of Physics,
Ankara University, Faculty of Sciences, 06100, Tando\u gan-Ankara,
Turkey\\}

\begin{abstract}
We show that higher degree Dirac currents of twistor and Killing
spinors correspond to the hidden symmetries of the background
spacetime which are generalizations of conformal Killing and Killing
vector fields respectively. They are the generalizations of 1-form
Dirac currents to higher degrees which are used in constructing the
bosonic supercharges in supergravity theories. In the case of
Killing spinors, we find that the equations satisfied by the higher
degree Dirac currents are related to Maxwell-like and
Duffin-Kemmer-Petiau equations. Correspondence between the Dirac
currents and harmonic forms for parallel and pure spinor cases is
determined. We also analyze the supergravity twistor and Killing
spinor cases in 10 and 11-dimensional supergravity theories and find
that although different inner product classes induce different
involutions on spinors, the higher degree Dirac currents still
correspond to the hidden symmetries of the spacetime.
\end{abstract}


\maketitle

\section{Introduction}

Twistor spinors and Killing spinors are defined as the solutions of
some differential equations on a spin manifold $M$ \cite{Baum
Friedrich Grunewald Kath,Baum,Leitner}. Killing spinors first arose
in mathematical physics to obtain the integrals of the geodesic
motion \cite{Hugston Penrose Sommers Walker,Chamseddine}. Later on,
it was found that Killing spinors are essential ingredients of
supersymmetry transformations in 10-dimensional and 11-dimensional
supergravity theories \cite{Duff Nilsson Pope1,Duff Nilsson Pope2}.
Killing spinors put some restrictions on the ambient manifold $M$
and hence the supergravity backgrounds depend on their existence and
their total number. The dimension of the space of Killing spinors
determine the number of supersymmetries in a bosonic supergravity
background. Killing vector fields and Killing spinors together
constitute a Killing superalgebra for the theory. In this
construction, 1-form Dirac currents and Lie derivatives of Killing
spinors are used \cite{Townsend,O Farril Jones Moutsopoulos,O
Farrill Papadopoulos,Alberca Lozano Ortin,van
Nieuwenhuizen,Gauntlett Pakis}. Therefore, investigating the
properties of manifolds that have Killing spinors is still a current
topic \cite{O Farrill Hustler,Harland Nolle}. Twistor spinors also
are used for different conceptual frameworks in physics literature
\cite{Penrose Rindler,Lichnerowicz1,Lichnerowicz2,Benn Kress1}. They
have conformal invariance property as opposed to the Killing spinors
and they contain Killing spinors as special cases, for this reason
they are also termed as conformal Killing spinors. Killing spinors
are twistor spinors that are at the same time eigenspinors of the
Dirac operator. Since the space of twistors contains Killing spinors
in a subspace, their generality ensures to take them as an initial
step towards obtaining supersymmetry generators in supergravity
theories. On the other hand, parallel spinors can also be defined as
special cases of Killing spinors and they are also used for the
construction of supersymmetry generators, because one can obtain
Killing spinors of a manifold $M$ from the parallel spinors of a
manifold $\tilde{M}$ by using the cone construction
\cite{Bar,Alekseevsky Cortes}.

Spinors can be defined as the elements of minimal left ideals of a
Clifford algebra. The spinor bilinears, constructed from a spinor
and an adjoint spinor via Clifford multiplication, are inhomogeneous
differential forms and can be resolved to a sum of their homogeneous
degree components by using the Fierz decomposition. So, one can use
projection operators to get a $p$-form field out of a spinor
bilinear. In this paper, we consider twistor spinors and Killing
spinors and use them to build up $p$-form Dirac currents with
investigating their special properties which are related to the
hidden symmetries of the background manifold. It is known that the
metric duals of 1-form Dirac currents of Killing spinors correspond
to Killing vector fields and they determine the bosonic supercharges
of supergravity theories. We show that $p$-form Dirac currents
correspond to the hidden symmetries of the background and they can
be used to define possible generalized supercharges in the theory.
By hidden symmetries we mean the generalizations of Killing and
conformal Killing vector fields. Twistor spinors can be used for
generating the conformal Killing-Yano (CKY) forms while Killing
spinors are related to the Killing-Yano (KY) forms of the background
in a similar way. Another result that appears when dealing with
twistor spinors is that the induced spinor bilinears satisfy an
equation generalizing the CKY equation to inhomogeneous forms.
Focusing on the Killing spinor case, different choices of spinor
inner products give rise to different closedness or co-closedness
properties of $p$-form Dirac currents. These properties are related
to Maxwell equations and Duffin-Kemmer-Petiau (DKP) equations. We
also analyze the $p$-form Dirac currents for the backgrounds of
11-dimensional supergravity and its Kaluza-Klein reduction, namely
10-dimensional supergravity. The existence of different bosonic
fields in supergravity theories affects the definitions of covariant
derivative and Dirac operator. This necessitates that the exterior
derivatives and co-derivatives of $p$-form Dirac currents contain
these bosonic supergravity fields. The choice of spinor inner
product classes also have an effect on $p$-form Dirac currents of
supergravity Killing spinors.

The organization of the paper is as follows. In Section 2, we define
the twistor spinors and Killing spinors with the differential
equations satisfied by them. Higher degree Dirac currents and their
properties are obtained in Section 3. It contains the hidden
symmetry properties of $p$-form Dirac currents of twistor and
Killing spinors and relations to Maxwell and DKP equations. In
Section 4, we consider the 11-dimensional and 10-dimensional
supergravity theories with properties of higher degree Dirac
currents on supergravity backgrounds. Section 5 concludes the paper.
There are also three appendices that contain the properties of
projection operators, spin structures on a manifold and the inner
product classes of spinors.

\section{Twistor and Killing Spinors}

Let $M$ be an $n$-dimensional Lorentzian spin manifold with metric tensor $g$, i.e., a spin structure
can be defined on its tangent bundle. The spinor fields are defined
as sections of the spinor bundle over $M$. Spinors can be seen as elements of the minimal left ideals of Clifford algebras (see Appendix B). A spin-invariant inner
product (spinor metric) can be defined in terms of adjoint
involutions on spinor fields. For two spinors $\psi$ and $\phi$, the
inner product $(\psi,\phi)$ is locally defined as follows
\begin{equation}
(\psi,\phi)=J^{-1}\psi^{\cal{J}}\phi
\end{equation}
where $J$ is a particular element of the Clifford algebra, ${\cal{J}}$ denotes the
adjoint involution which can be taken as $\xi$ or $\xi\eta$ (see
footnotes [36], [37] and Appendix C) and also (in the complex case) their
compositions with complex conjugation. Here juxtaposition corresponds to the Clifford product. The Clifford product of a 1-form $A$ and an arbitrary form $\Phi$ is given as follows
\begin{eqnarray}
A\Phi&=&A\wedge\Phi+i_{\tilde{A}}\Phi\nonumber\\
\Phi A&=&A\wedge\Phi^{\eta}-i_{\tilde{A}}\Phi^{\eta}\nonumber
\end{eqnarray}
where $\wedge$ denotes the wedge product, the vector field $\tilde{A}$ corresponds to the metric dual of $A$ and $i_{\tilde{A}}$ is the interior derivative with respect to $\tilde{A}$. The dual of a spinor is
defined from the definition of the inner product as
\begin{equation}
\bar{\psi}=J^{-1}\psi^{\cal{J}}
\end{equation}
which lies in the corresponding minimal right ideal of the algebra and note that this operation is general than the Dirac adjoint operation defined in the literature. The spinor covariant derivative with respect to a vector field $X$,
which is denoted by $\nabla_X$, can be used to build a first-order
elliptic differential operator called the Dirac operator on spinor
fields. It is defined as the Clifford product of the co-frame basis elements
$e^a$ (not necessarily $g$-orthonormal) and the spinor covariant derivative with respect to the frame basis;
\begin{equation}
\displaystyle{\not}D=e^a\nabla_{X_a}.
\end{equation}
The eigenspinors
of the Dirac operator, namely the complex spinors that satisfy the Dirac
equation, are called Dirac spinors;
\begin{equation}
\displaystyle{\not}D\psi=m\psi
\end{equation}
where $m$ is a constant. Another first-order operator which is
defined on spinor fields is the twistor (Penrose) operator and it is
written in terms of spinor covariant derivative and the Dirac
operator as follows
\begin{equation}
{\cal{P}}_X:=\nabla_X-\frac{1}{n}\tilde{X}\displaystyle{\not}D
\end{equation}
where $\tilde{X}$ is the 1-form that is the metric dual of the
vector field $X$. The logic lying under the construction of this
operator is based on projecting the covariant derivatives of spinors
onto the kernel of left Clifford multiplication so
${\cal{P}}:= pr\circ \nabla $, here $pr$ is the corresponding projection.
The spinors that are in the kernel of the twistor operator, namely
those that satisfy the following equation are called twistor spinors;
\begin{equation}
\nabla_X\psi=\frac{1}{n}\tilde{X}\displaystyle{\not}D\psi.
\end{equation}
Twistor spinors which are at the same time Dirac spinors are
defined as Killing spinors and they are solutions of the following
equation
\begin{equation}
\nabla_X\psi=\lambda\tilde{X}\psi
\end{equation}
where $\lambda$ is a complex constant and is called as the \emph{Killing number}. In
fact, the existence of Killing spinors confines the geometry of $M$. If $R(X,Y):=[\nabla_X,\nabla_Y]-\nabla_{[X,Y]}$ is the curvature operator of the spinor connection; then it is known that its effect on an arbitrary spinor field can be written as the left Clifford action of the curvature $2$-forms ${R^a}_b$ of torsion-free connection and is given by
\begin{equation}
R(X,Y)\psi=\frac{1}{2}e^a(X)e^b(Y)R_{ab}\psi\qquad\forall X,Y \in\Gamma(TM);\, \forall \psi \in\Gamma({\cal S}(M))
\end{equation}
where $\Gamma(TM)$ and $\Gamma({\cal S}(M))$ denote the smooth sections of the tangent bundle and spinor bundle respectively and $\{e^a\}$ is a $g$-orthonormal co-frame basis. From now on orthonormality condition for local co-frames is assumed unless otherwise stated (see Appendix A). For $\psi$ a Killing spinor with Killing number $\lambda$ then by using the torsion zero condition
\begin{equation}
R(X,Y)\psi=-\lambda^2[\widetilde{X},\widetilde{Y}]_{_{Cl.}}\psi
\end{equation}
where $[.,.]_{_{Cl.}}$ is the Clifford commutator. By equating the right hand sides of the above equations one can find
\begin{equation}
R_{ab}\psi=-4\lambda^2 e_{ab}\psi
\end{equation}
and also using the torsion-free Bianchi identities in the language of Clifford calculus one gets
\begin{equation}
P_{b}\psi=-4\lambda^2 (n-1)e_{b}\psi
\end{equation}
with $P_{b}$'s denoting the Ricci $1$-forms. In terms of the cotangent bundle endomorphisms $\widehat{Ric}$ and $Id_{T\!^{^{*}}\!M}$ where $\widehat{Ric}(e_{b}):=P_{b}$ this equation can be written as
\begin{equation}
\left(\widehat{Ric}(e_{b})+4\lambda^2 (n-1)Id_{T\!^{^{*}}\!M}(e_{b})\right)\psi=0\qquad\forall b
\end{equation}
which means that the image of the endomorphism $\widehat{Ric}+4\lambda^2 (n-1)Id_{T\!^{^{*}}\!M}$ is totally null. And by Clifford multiplying the both sides of (11) with $e^b$ from the left and by using $e^b\,e_b=n$ we get the scalar curvature as
\begin{equation}
{\cal{R}}=-4\lambda^2 n(n-1)
\end{equation}
which is constant and implies that the Killing number must be real or pure imaginary.

A particular subspace of Killing spinors which have Killing number
$\lambda=0$ are called parallel spinors and satisfy the following
equation;
\begin{equation}
 \nabla_X\psi=0  .
\end{equation}
Parallel spinors and Killing spinors are also related to each other
in a different manner: Killing spinors of $M$ can be
obtained from parallel spinors of another manifold $\tilde{M}$ by
using the cone construction method. Moreover, the existence of parallel spinors restricts the holonomy group of the manifold.

\section{$p$-form Dirac Currents}

Let us denote the spinor space by ${\cal S}$ and its dual by ${\cal S}^{*}$ which are defined respectively as minimal left and right ideals of a Clifford algebra $Cl_{r,s}$ with $r$ positive and $s$ negative generators. The tensor product of ${\cal S}$ with ${\cal S}^{*}$ gives the algebra of endomorphisms over the spinor
space $End\,{\cal S} \simeq {\cal S}\otimes {\cal S}^{*}$ and
this algebra is either the Clifford algebra of convenient dimension
or a subalgebra of it, and the corresponding algebra isomorphisms are
known as the Fierz Identities. Identifying a minimal left
ideal of the Clifford algebra with the spinor space gives the
advantage of multiplying spinors and/or dual spinors by using the
Clifford product since all ideals are also subalgebras. So, if
$\psi$ is a spinor and $\overline{\phi}$ a dual spinor then by Fierz
isomorphism one can write $\psi\otimes\overline{\phi}=\psi \overline{\phi}$. The
Clifford product of a spinor $\psi$ and an adjoint spinor
$\overline{\phi}$ can be written as an inhomogeneous
differential form in terms of projectors $\wp_p$ on $p$-form components. The related $0$-form components can be transformed into inner products of the spinors under consideration where $\overline{\phi}\psi=(\phi,\psi)$ and $(.\,,.)$ has one of the subgroups of the Clifford group as the isometry group (usually taken as the Spin group). So
\begin{eqnarray}
\psi\bar{\phi}&=&\sum_{p=0}^n\wp_{_{p}}(\psi\bar{\phi})=\sum_{p=0}^n\wp_{_{0}}(\psi\bar{\phi}\,e_{I(p)}^{\xi})e^{I(p)}=\sum_{p=0}^n\wp_{_{0}}(\bar{\phi}\,e_{I(p)}^{\xi}\psi)e^{I(p)}=\sum_{p=0}^n(\phi,\,e_{I(p)}^{\xi}\psi)e^{I(p)}\\
&=&(\phi,\psi)+(\phi,e_a\psi)e^a+(\phi,e_{ba}\psi)e^{ab}+...+(\phi,e_{a_p...a_2\,a_1}\psi)e^{a_1
a_2...a_p}+...+(-1)^{\lfloor n/2\rfloor} (\phi,z\psi)z\nonumber
\end{eqnarray}
where $e^{a_1a_2...a_p}:=e^{a_1}\wedge e^{a_2}\wedge...\wedge
e^{a_p}=e^{a_1} e^{a_2}... e^{a_p}$ and $z=*1$ is the
globally defined volume form (when $M$ is orientable), $*$ denotes the Hodge map and multi-index $I(p)$ is a well-ordered $p$-tuple of indices. Hence, one can obtain different degree $p$-forms from two spinors by using
projection operators on different form degrees (see Appendix A).
For example, the 1-form component of the Clifford product
$\psi\bar{\phi}$ is denoted as
\begin{equation}
j^{(1)}_{\psi,\phi}=\wp_{_1}(\psi\bar{\phi})=(\phi,e_a\psi)\,e^a.
\end{equation}
Generally a $p$-form component is written as
\begin{equation}
\wp_p(\psi\bar{\phi})=(\phi,e_{a_p...a_2\,a_1}\psi)e^{a_1a_2...a_p}.
\end{equation}
The reality conditions of Dirac currents depend on their degrees and
the types of the chosen inner product (or equivalently the adjoint
involutions). These conditions are only apt to
$\mathbb{C}^*$-symmetric inner product classes, because other
classes are not affected by the operation of complex conjugation.
The Dirac currents with adjoint involutions ${\cal{J}}\simeq\xi\eta$
or $\xi\eta^*$ are defined by

\begin{eqnarray}
j^{(p)}_{\psi}=\sqrt{(-1)^{\lfloor
p/2\rfloor}}\,\wp_p(\psi\bar{\psi}) &=&\left\{
                                                                               \begin{array}{ll}
                                                                                 \wp_p($i$\psi\bar{\psi}), & \hbox{$\lfloor p/2 \rfloor$ and $p$ have different parities ;} \\
                                                                                 \wp_p(\psi\bar{\psi}), & \hbox{$\lfloor p/2 \rfloor$ and $p$ have same parity.}
                                                                               \end{array}
                                                                             \right.
\end{eqnarray}
and the remaining ones for which ${\cal{J}}\simeq\xi$ or $\xi^*$ are
determined only by the parity of ${\lfloor p/2\rfloor}$ where odd
${\lfloor p/2\rfloor}$ gives $\wp_p($i$\psi\bar{\psi})$ and even
${\lfloor p/2\rfloor}$ gives $\wp_p(\psi\bar{\psi})$. So all
$j^{(p)}_{\psi}$'s are real $p$-form fields and the i used above is
the complex imaginary unit. For other inner product classes
\cite{Benn Tucker} the Dirac current is equal to
$\wp_p(\psi\bar{\psi})$ since it is real in those cases.

Let us denote the spinor bilinear of $\psi$ as
$a_{\psi}:=\psi\bar{\psi}$. Since $j^{(p)}_{\psi}$ and
$\wp_p(a_{\psi})$ are identical or $\pm$i times of each other, we
will make use of $\wp_p(a_{\psi})$ as the $p$-form Dirac current
instead of $j^{(p)}_{\psi}$ unless an operation that violates
$\mathbb{C}$-linearity occurs. These $p$-form Dirac currents namely
the homogeneous components of spinor bilinears can be used to
classify spinor fields in various dimensions \cite{Bonora Brito
Rocha} and they also give a unique criterion for a spinor to be pure
\cite{Charlton}. When $\psi$ is a Killing spinor or a twistor
spinor, the special case of 1-form Dirac current $j^{(1)}_{\psi}$
correspond to the metric dual of a Killing vector field or a
conformal Killing vector field of the background manifold
respectively \cite{Kuhnel Rademacher}.

\subsection{Twistor spinor case}

If $\psi$ is a twistor spinor, then as it is shown below each of the Dirac currents of
$\psi$ are CKY forms. A $p$-form
$\omega_{(p)}$ is a CKY $p$-form, if it satisfies the following
equation for every vector field $X$ in $n$ dimensions;
\begin{equation}
\nabla_{X}\omega_{(p)}=\frac{1}{p+1}i_{X}d\omega_{(p)}-\frac{1}{n-p+1}\widetilde{X}\wedge\delta\omega_{(p)}
\end{equation}
where $d$ is the exterior derivative and $\delta$ is the co-derivative operator.

By taking the covariant derivative of the $p$-form Dirac current
$\wp_p(a_{\psi})$ of a twistor spinor $\psi$ with respect to the frame field $X_b$, one obtains
\begin{eqnarray}
\nabla_{X_b}\wp_p(a_{\psi})&=&\wp_p\left((\nabla_{X_b}\psi)\bar{\psi}+\psi(\nabla_{X_b}\bar{\psi})\right)\nonumber\\
&=&\frac{1}{n}\wp_p\left((e_b\displaystyle{\not}D\psi)\bar{\psi}+\psi\overline{\displaystyle{\not}D\psi}(e_b)^{\cal{J}}\right)\nonumber\\
&=&\frac{1}{n}\wp_p(e_b\,e^a(\nabla_{X_a}\psi)\bar{\psi}+\psi\overline{\nabla_{X_a}\psi}e^a\,e_b)
\end{eqnarray}
where we used the twistor equation (6), the compatibility of $\nabla_X$ with the inner product $(. ,.)$, namely $\nabla_X\bar{\psi}=\overline{\nabla_X\psi}$ and $(e_b\,e^a)^{\cal{J}}=e^a\,e_b$. An important thing here is that the occurance of the spinor inner product ascribed to ${\cal{J}}$ does not change the calculations since the terms where it appears does not depend on a particular choice of the inner product for the twistor spinor case. However, for Killing spinors the situation will be different. Using the identities
\begin{eqnarray}
(\nabla_{X_a}\psi)\bar{\psi}&=&\nabla_{X_a}(\psi\bar{\psi})-\psi\overline{\nabla_{X_a}\psi}\\
e_b\,e^a&=&{\delta_b}^a+e_b\wedge e^a\\
\wp_p(e_b\displaystyle{\not}da_{\psi})&=&e_b\wedge\wp_{p-1}(\displaystyle{\not}da_{\psi})+i_{X_b}\wp_{p+1}(\displaystyle{\not}da_{\psi})
\end{eqnarray}
and after doing some manipulations, the covariant derivative of the $p$-form
Dirac current is found as
\begin{eqnarray}
n\nabla_{X_b}\wp_p(a_{\psi})&=&e_b\wedge\wp_{p-1}(\displaystyle{\not}da_{\psi}-2e^a\wedge (b_{a})_{\psi})+i_{X_b}\wp_{p+1}(\displaystyle{\not}da_{\psi}-2i_{X^a}(b_{a})_{\psi})
\end{eqnarray}
where $\displaystyle{\not}d=d-\delta$ is the Hodge-de Rham operator and $(b_{a})_{\psi}:=\psi\overline{\nabla_{X_a}\psi}=a_{\psi,\nabla_{X_a}\psi}$. By taking wedge
product with $e^b$ and interior derivative $i_{X^b}$ of (24) and also using the property $e^b\wedge i_{X^b} \omega_{(p)}=p\,\omega_{(p)}$
for a $p$-form $\omega_{(p)}$, the following two identities can be obtained respectively
\begin{eqnarray}
nd\wp_p(a_{\psi})&=&(p+1)\wp_{p+1}(\displaystyle{\not}da_{\psi}-2i_{X^a}(b_{a})_{\psi})\\
n\delta\wp_p(a_{\psi})&=&(-n+p-1)\wp_{p-1}(\displaystyle{\not}da_{\psi}-2e^a\wedge (b_{a})_{\psi}).
\end{eqnarray}
Using (25) and (26) in (24) we reach to the CKY equation for
$p$-form Dirac currents
\begin{eqnarray}
\nabla_{X_b}\wp_p(a_{\psi})=\frac{1}{p+1}i_{X_b}d\wp_p(a_{\psi})-\frac{1}{n-p+1}e_b\wedge\delta\wp_p(a_{\psi}).
\end{eqnarray}
By using (A7) this equation can be rewritten in a more compact form
as
\begin{eqnarray}
\nabla_{X_b}\wp_p(a_{\psi})=\left[\frac{1}{p+1}\,i_{X_b}d+\frac{1}{n-p+1}\,*^{-1}i_{X_b}d*\right] \wp_p(a_{\psi})
\end{eqnarray}
and since the operator $i_{X}d$ and its Hodge conjugate
$*^{-1}i_{X}d*$ are degree preserving operations, they
commute with projectors $\wp_p$. Summing over $p$ and using (A2),
an inhomogeneous equation can be found as
\begin{eqnarray}
\nabla_{X_b}a_{\psi}=\left[\left(\sum_{p=0}^n\frac{\wp_{_{p}}}{p+1}\right)\,i_{X_b}d+\left(\sum_{p=0}^n\frac{\wp_{_{p}}}{n-p+1}\right)\,*^{-1}i_{X_b}d*\right] a_{\psi}
\end{eqnarray}
for the spinor bilinear $a_{\psi}$. This equation generalizes the CKY equation to inhomogeneous forms, particularly to spinor bilinears. Since the spinor bilinears are constructed from physical quantities that are spinor fields, the CKY equation which has solutions corresponding to the geometrical symmetries of the background acquires a physical meaning by this way.

On the other hand, if we define $\wp:=\left(\sum_{p=0}^n\frac{\wp_{_{p}}}{p+1}\right)$ and
$\widehat{\wp}:=\left(\sum_{p=0}^n\frac{\wp_{_{p}}}{n-p+1}\right)$
then from (A.4) and the equality
$\sum_{p=0}^n\frac{\wp_{_{p}}}{p+1}=\sum_{p=0}^n\frac{\wp_{_{n-p}}}{n-p+1}$
we see that $\widehat{\wp}=\,*^{-1}\wp*$ which implies that the last equation above
turns into
\begin{eqnarray}
\nabla_{X}a_{\psi}=\left[\wp\,i_{X}d+\,*^{-1}(\wp\,i_{X}d)*\right] a_{\psi}.
\end{eqnarray}
which is a Hodge covariant equation, i.e.
\begin{eqnarray}
\nabla_{X} *a_{\psi}=\left[*^{-1}(\wp\,i_{X}d)*+\wp\,i_{X}d\right]*a_{\psi}.
\end{eqnarray}
Here $X$ is an arbitrary vector field and the equality $\,*(\wp\,i_{X}d)*^{-1}=\,*^{-1}(\wp\,i_{X}d)*$ is used which is valid for all homogeneous linear operators of degree zero. By using the property
$(p+1)\wp_{_{p}}\wp=\wp_{_{p}}$ obtained from (A.2) and (A3) one can
turn back to the homogeneous equation (CKY equation for a $p$-form)
(28).

In fact, CKY forms are generalizations of conformal Killing vector
fields to higher degree forms and are used to generate the symmetry
operators of massless Dirac spinors \cite{Benn Charlton}. As we have seen, every
twistor spinor defines a CKY $p$-form through its $p$-form Dirac
current and this correspondence is relevant also in the Riemannian backgrounds \cite{Semmelmann}. This means that twistor spinors can be used to generate the symmetries of massless Dirac spinors.

\subsection{Killing spinor case}

For a Killing spinor $\psi$, the covariant derivative of $p$-form Dirac currents is highly dependent on the choice of the spinor inner product. However, we show that in all choices of the inner product, $p$-form Dirac currents correspond to KY forms. A KY form
$\omega_{(p)}$ is a co-closed CKY form $\delta\omega_{(p)}=0$ which
satisfy the equation
\begin{equation}
\nabla_{X_a}\omega_{(p)}=\frac{1}{p+1}i_{X_a}d\omega_{(p)}.
\end{equation}
The covariant derivative of $\wp_p(a_{\psi})$ with respect to $X_b$ is written as
\begin{equation}
\nabla_{X_b}\wp_p(a_{\psi})=\wp_p((\nabla_{X_b}\psi)\bar{\psi}+\psi \nabla_{X_b}\bar{\psi}).
\end{equation}
From the compatibility of $\nabla_X$ with the inner product $(.\,,.)$ and the Killing spinor equation (7), one can write
\begin{equation}
\nabla_{X_b}\bar{\psi}=\overline{\nabla_{X_b}\psi}=\lambda^j\bar{\psi}(e_b)^{\cal{J}}.
\end{equation}
Here $j$ can be chosen as the identity (Id) or the complex conjugation $(^*)$ and ${\cal{J}}$ can be chosen as the involutions $\xi$, $\xi\eta$, $\xi^*$ or $\xi\eta^*$ (see Appendix C). Since $e^a$ are real 1-forms the involutions $\xi^*$ and $\xi\eta^*$ does not give different results from $\xi$ and $\xi\eta$. We will denote the chosen $j$ and ${\cal{J}}$ as a pair $\{j,{\cal{J}}\}$. These different choices give
\begin{equation}
\lambda^j\bar{\psi}(e_b)^{\cal{J}}=\left\{
                                     \begin{array}{ll}
                                       \lambda\bar{\psi}e_b, & \hbox{for $\{Id , \xi\}$ , $\{Id , \xi^*\}$;} \\
\\
                                       -\lambda\bar{\psi}e_b, & \hbox{for $\{Id , \xi\eta\}$ , $\{Id , \xi\eta^*\}$;} \\
\\
                                       \lambda^*\bar{\psi}e_b, & \hbox{for $\{^* , \xi\}$ , $\{^* , \xi^*\}$;} \\
\\
                                       -\lambda^*\bar{\psi}e_b, & \hbox{for $\{^* , \xi\eta\}$ , $\{^* , \xi\eta^*\}$.}
                                     \end{array}
                                   \right
.
\end{equation}
From (33), the covariant derivative of $p$-form Dirac currents can be calculated as follows
\begin{eqnarray}
\nabla_{X_b}\wp_p(a_{\psi})&=&\wp_p\left(\lambda e_ba_{\psi}+\lambda^ja_{\psi}(e_b)^{\cal{J}}\right)\nonumber\\
&=&\wp_p(\lambda e_ba_{\psi}-\lambda^j(e_b)^{\cal{J}}(a_{\psi})^{\eta})+2\lambda^j(e_b)^{\cal{J}}\wedge \wp_{p-1}((a_{\psi})^{\eta})\nonumber
\end{eqnarray}
where we used $a_{\psi}(e_b)^{\cal{J}}=-(e_b)^{\cal{J}}(a_{\psi})^{\eta}+2e_b\wedge(a_{\psi})^{\eta}$. By defining the operator $\lambda_{(p)}^{\pm}:=(\lambda\pm(-1)^p\lambda^j{\cal{J}})$ the last equation can be written as follows
\begin{eqnarray}
\nabla_{X_b}\wp_p(a_{\psi})=\left(\lambda_{(p)}^{-} e_b\right)\wedge\wp_{p-1}(a_{\psi})+i_{\widetilde{\lambda_{(p)} ^{+}e_b}}\wp_{p+1}(a_{\psi}).
\end{eqnarray}

Bearing in mind that the connection $\nabla$ is torsion-free, then the following identities for exterior derivative and
co-derivative operations are obtained respectively as
\begin{eqnarray}
d\wp_p(a_{\psi})=\left(e^b\lambda_{(p)} ^{+}e_b\right) \frac{p+1}{n}\wp_{p+1}(a_{\psi}),
\end{eqnarray}

\begin{eqnarray}
\delta\wp_p(a_{\psi})=-\left(e^b\lambda_{(p)} ^{-}e_b\right) \frac{n-p+1}{n}\wp_{p-1}(a_{\psi}).
\end{eqnarray}
The explicit forms of equations (37) and (38) for each choice of $\{j , \cal{J}\}$ can be found by using the Table 1 and since $n^{-1}\left(e^a\lambda_{(p)} ^{\pm}e_a\right)e_b=\lambda_{(p)} ^{\pm}e_b $ one can also extract (36) for each case.
\begin{table}[h]
\centering
\begin{tabular}{c|c}

  $ \{j , \cal{J}\}$ & $\left(e^b\lambda_{(p)} ^{\pm}e_b\right) / n $  \\ \hline
  $\{Id , \xi\}$ , $\{Id , \xi^*\}$ & $\lambda\pm(-1)^p\lambda$ \\
  $\{Id , \xi\eta\}$ , $\{Id , \xi\eta^*\}$ & $\lambda\mp(-1)^p\lambda$ \\
  $\{^* , \xi\}$ , $\{^* , \xi^*\}$ & $\lambda\pm(-1)^p\lambda^*$ \\
  $\{^* , \xi\eta\}$ , $\{^* , \xi\eta^*\}$ & $\lambda\mp(-1)^p\lambda^*$ \\

\end{tabular}
\caption{Values of $(e^b\lambda_{(p)} ^{\pm}e_b) / n$ for different involutions.}
\end{table}

Since $\lambda$ is real ($\lambda=\lambda^*$) or pure imaginary ($\lambda=-\lambda^*$), the different cases for even or odd $p$ degrees can be analyzed. The reality condition of Dirac currents given by (18) constrains the considered possibilities through the compatibility of the equations (37) and (38) as follows: These equations contain the degree $(p-1)$,$p$ and $(p+1)$-form components of the spinor bilinear $a_{\psi}$, since (18) is used, the floor functions of the half of these degrees are needed and are given in Table 2.
\begin{table}[h]
\centering
\begin{tabular}{c|c|c|c}

  $p$ & $\lfloor \frac{p-1}{2}\rfloor$ & $\lfloor\frac{p}{2}\rfloor$ & $\lfloor\frac{p+1}{2}\rfloor$  \\ \hline
  $2k$ & $k-1$ & $k$ & $k$ \\
  $2k+1$ & $k$ & $k$ & $k+1$ \\

\end{tabular}
\caption{Floor functions of $(p-1)/2$, $p/2$ and $(p+1)/2$ for even and odd $p$-form degrees.}
\end{table}

Finally the neat relation between the Dirac currents and the homogeneous components of the spinor bilinear for ${\cal{J}}=\xi$ or $\xi^*$ can be seen from Table 3.

\begin{table}[h]
\centering
\begin{tabular}{c|c|c|c|c}

  $p$ &$k$ & $j^{(p-1)}_{\psi}$ & $j^{(p)}_{\psi}$ & $j^{(p+1)}_{\psi}$  \\ \hline
  $2k$ &$$even$$ &$\wp_{p-1}($i$a_{\psi})$ & $\wp_p(a_{\psi})$ & $\wp_{p+1}(a_{\psi})$  \\
  $2k$ &$$odd$$ & $\wp_{p-1}(a_{\psi})$ & $\wp_p($i$a_{\psi})$ & $\wp_{p+1}($i$a_{\psi})$  \\ \hline                                                                              $2k+1$ &$$even$$ & $\wp_{p-1}(a_{\psi})$ & $\wp_p(a_{\psi})$ & $\wp_{p+1}($i$a_{\psi})$  \\
  $2k+1$ &$$odd$$ & $\wp_{p-1}($i$a_{\psi})$ & $\wp_p($i$a_{\psi})$ & $\wp_{p+1}(a_{\psi})$  \\

\end{tabular}
\caption{Relation between the Dirac currents and the homogeneous components of the spinor bilinear for different $p$-form degrees in the case of ${\cal{J}}=\xi$ or $\xi^*$.}
\end{table}

In the case of ${\cal{J}}=\xi\eta$ or $\xi\eta^*$ the columns $j^{(p-1)}_{\psi}$ and $j^{(p+1)}_{\psi}$ switch their places and $j^{(p)}_{\psi}$ column remains in the same place. As a result, there are eight cases that are compatible with the reality conditions. We have four different cases that correspond to closed $p$-form and co-exact $(p-1)$-form Dirac currents which are\\
(i) $\lambda=\lambda^*$, $p$ odd, $\{Id , \xi\}$, $\{Id , \xi^*\}$, $\{^* , \xi\}$ or $\{^* , \xi^*\}$,\\
(ii) $\lambda=-\lambda^*$, $p$ even, $\{^* , \xi\}$ or $\{^* , \xi^*\}$,\\
(iii) $\lambda=\lambda^*$, $p$ even, $\{Id , \xi\eta\}$, $\{Id , \xi\eta^*\}$, $\{^* , \xi\eta\}$ or $\{^* , \xi\eta^*\}$,\\
(iv) $\lambda=-\lambda^*$, $p$ odd, $\{^* , \xi\eta\}$ or $\{^* , \xi\eta^*\}$,
\begin{eqnarray}
d\wp_p(a_{\psi})&=&0\nonumber\\
\delta\wp_p(a_{\psi})&=&-2\lambda(n-p+1)\wp_{p-1}(a_{\psi})
\end{eqnarray}
and also four different cases for co-closed $p$-form and exact $(p+1)$-form Dirac currents which are\\
(i') $\lambda=\lambda^*$, $p$ even, $\{Id , \xi\}$, $\{Id , \xi^*\}$, $\{^* , \xi\}$ or $\{^* , \xi^*\}$,\\
(ii') $\lambda=-\lambda^*$, $p$ odd, $\{^* , \xi\}$ or $\{^* , \xi^*\}$,\\
(iii') $\lambda=\lambda^*$, $p$ odd, $\{Id , \xi\eta\}$, $\{Id , \xi\eta^*\}$, $\{^* , \xi\eta\}$ or $\{^* , \xi\eta^*\}$,\\
(iv') $\lambda=-\lambda^*$, $p$ even, $\{^* , \xi\eta\}$, $\{^* , \xi\eta^*\}$,
\begin{eqnarray}
d\wp_p(a_{\psi})&=&2\lambda(p+1)\wp_{p+1}(a_{\psi})\nonumber\\
\delta\wp_p(a_{\psi})&=&0.
\end{eqnarray}

In the last four cases, the $p$-form Dirac currents are co-closed and by comparing (36) with (40), one obtains that they satisfy the KY equation (32)
\begin{eqnarray}
\nabla_{X_b}\wp_p(a_{\psi})=\frac{1}{p+1}i_{X_b}d\wp_p(a_{\psi})\nonumber.
\end{eqnarray}
In the first four cases, the $p$-form Dirac currents are closed and they satisfy the $(n-p)$-form KY equation for the Hodge duals of the $p$-form Dirac currents
\begin{eqnarray}
\nabla_{X_b}*\wp_p (a_{\psi})=\frac{1}{n-p+1}i_{X_b}d*\wp_p(a_{\psi})\nonumber.
\end{eqnarray}

KY forms are generalizations of Killing vector fields to higher
degree forms and are generators of the symmetry operators for
massive Dirac spinors \cite{Benn Kress2,Acik Ertem Onder Vercin1}. As a result, the Killing spinors whose $p$-form Dirac
currents satisfy the KY equation generate the symmetries of massive
Dirac spinors.

Closedness and co-closedness of Dirac currents represent their
conservational properties. In the closed cases $d\wp_p(a_{\psi})=0$
which are the first four cases in (39), the $p$-form Dirac currents
$\wp_p(a_{\psi})$ themselves correspond to conserved currents, while
in the co-closed cases $\delta\wp_p(a_{\psi})=0$ which are the
second four cases in (40), the Hodge duals of $p$-form Dirac currents
$*\wp_p(a_{\psi})$ correspond to conserved currents. In this way,
the 1-form Dirac currents that correspond to Killing vector fields,
are used to construct bosonic supercharges in supergravity theories.
Hence, $p$-form Dirac currents of Killing spinors which are KY forms
of the background can be used to define generalized bosonic
supercharges in supergravity.

The equations satisfied by $p$-form Dirac currents in closed and
co-exact cases have an analogous structure with Maxwell equations.
In these cases, we have
\begin{eqnarray}
d*\wp_p(a_{\psi})&=&-2(-1)^p\lambda(n-p+1)*\wp_{p-1}(a_{\psi})\nonumber\\
d\wp_p(a_{\psi})&=&0.
\end{eqnarray}
Hence, Dirac currents $\wp_p(a_{\psi})$ behave like the field
strength of the Maxwell field and the source term is constructed
from the Hodge duals of the one lower degree Dirac currents. In
co-closed and exact cases, a similiar structure appears for Hodge
duals of Dirac currents and we have
\begin{eqnarray}
d\wp_p(a_{\psi})&=&2\lambda(p+1)\wp_{p+1}(a_{\psi})\nonumber\\
d*\wp_p(a_{\psi})&=&0.
\end{eqnarray}
So, $*\wp_p(a_{\psi})$ behave like the Maxwell field strength and
source term is the Dirac current of one higher degree. The duality
between closed and co-closed cases correspond to a duality of field
equations and Bianchi identities through $p$-form Dirac currents.
This means that Noether-like charges constructed from $p$-form Dirac
currents in one case will have a dual topological charge constructed
from $p$-form Dirac currents in the other case, because they are
Hodge duals of each other.

In fact, these set of equations have also interesting relations with
Duffin-Kemmer-Petiau (DKP) equations \cite{Lichnerowicz3}. Let us consider a $p$-form Dirac
current $j^{(p)}_{\psi}$ that satisfy (39), then
$j^{(p-1)}_{\psi}$ must satisfy (40). Hence, we have the set
of equations
\begin{eqnarray}
dj^{(p-1)}_{\psi}=2\lambda pj^{(p)}_{\psi}\quad&,&\quad\delta j^{(p-1)}_{\psi}=0\\
\delta j^{(p)}_{\psi}=-2\lambda(n-p+1)j^{(p-1)}_{\psi}\quad&,&\quad
dj^{(p)}_{\psi}=0.
\end{eqnarray}
If we choose $n=2p-1$, then we obtain the following
DKP equations
\begin{eqnarray}
d\Phi_\pm=\mu\Phi_\mp\quad&,&\quad\delta\Phi_\pm=0\\
\delta\Phi_\mp=-\mu\Phi_\pm\quad&,&\quad d\Phi_\mp=0
\end{eqnarray}
where $\Phi_\pm=j^{(p-1)}_{\psi}$, $\Phi_\mp=j^{(p)}_{\psi}$ for $p$
odd or even and $\mu=2\lambda p$ which corresponds to the mass and
it is always real because of the analysis led to (39) and (40). This
implies that in odd dimensions $n=2p-1$, the $p$-form and
$(p-1)$-form Dirac currents are a couple of bosonic fields that
satisfy the DKP equations. $p$-form Dirac currents can be related to
bosonic supercharges on $(p-1)$-branes and this means that charges
of $(p-1)$-branes and $(p-2)$-branes are coupled together through
DKP equations. In this particular dimension if $p$-form and
$(p-1)$-form Dirac currents satisfy
$Q:=*j^{(p)}_{\psi}=j^{(p-1)}_{\psi}$, then the Maxwell-like
equations (41) assume a very special form that is $dQ=*Q$.

On the other hand, by combining the effects of exterior derivative and co-derivative operators on $p$-form Dirac currents, one can obtain
the effect of Laplace-Beltrami operator $\Delta=-d\delta-\delta d$
on them. In all cases, they correspond to the eigenforms of the Laplace-Beltrami operator. However, the eigenvalues are different for different cases. For the first four cases (39), we obtain
\begin{equation}
\Delta\wp_p(a_{\psi})=4\lambda^2p(n-p+1)\wp_p(a_{\psi})
\end{equation}
and for the last four cases (40)
\begin{equation}
\Delta\wp_p(a_{\psi})=4\lambda^2(p+1)(n-p+2)\wp_p(a_{\psi}).
\end{equation}

\subsection{Parallel and pure spinor cases}

If $\psi$ is a parallel spinor, then as can be seen from (21) that
the $p$-form Dirac currents satisfy the equation
\begin{equation}
\nabla_{X}\wp_p(a_{\psi})=0
\end{equation}
and hence they are parallel forms. This implies that the $p$-form
Dirac currents of parallel spinors are harmonic forms, because thay
are both closed and co-closed and satisfy the following
equality
\begin{equation}
\Delta\wp_p(a_{\psi})=0
\end{equation}
and the last two equations are both valid for the inhomogeneous spinor bilinear $a_{\psi}$ itself because of (A2).

If $\psi$ is a Killing spinor and is a pure spinor at the same time,
then the $p$-form Dirac currents again correspond to harmonic forms.
In even dimensions $n=2r$ with maximal Witt index, a spinor $\psi$
is a pure spinor if and only if $\wp_p(a_{\psi})=0$ for $p\neq r$
\cite{Charlton}. Hence, only the middle form Dirac current is
non-zero for a pure spinor. In the Killing spinor case, the exterior
derivative and co-derivative of a $p$-form Dirac current includes
$(p+1)$-form or $(p-1)$-form Dirac currents and this implies that
for pure spinors, the middle form Dirac current is both closed and
co-closed which makes it a harmonic form.

\section{Supergravity Dirac Currents}

In supergravity theories, Killing spinors are important in defining
the generators of supersymmetry algebra. The number of Killing
spinors correspond to the number of supersymmetry generators in a
supergravity background. However, the spinor covariant derivative
and hence the Dirac operator are modified by the additional
supergravity fields in a supergravity background. So, one must
consider the modified covariant derivatives and Dirac operators to define the Killing and
twistor spinors. As a result of this, the properties of Dirac currents are
also modified by the supergravity fields.

As an example, we consider the 11-dimensional supergravity
background which is a triple ($M,g,F$), where $M$ is an
11-dimensional Lorentzian spin manifold with metric $g$ and $F$ is a
closed 4-form which is the field strength of the bosonic 3-form
field $A$. The bosonic part of the 11-dimensional supergravity
action is
\begin{equation}
S=\frac{1}{2\kappa_{11}^2}\int\left(R_{ab}\wedge
*e^{ab}-\frac{1}{2}F\wedge*F-\frac{1}{6}A\wedge F\wedge F\right)
\end{equation}
where $\kappa_{11}$ denotes the 11-dimensional gravitational coupling
constant. The field equations of this theory are given by
\begin{eqnarray}
d*F&=&\frac{1}{2}F\wedge F\\
Ric(X,Y)*1&=&\frac{1}{2}i_XF\wedge*i_YF-\frac{1}{6}g(X,Y)F\wedge*F
\end{eqnarray}
where $Ric$ is the Ricci tensor.

The spinor covariant derivative in the supergravity background is
defined in terms of the ordinary spinor covariant derivative $\nabla_X$
and the closed 4-form field $F$ as follows \cite{O Farrill Meessen Philip}
\begin{equation}
{\cal{D}}_X=\nabla_X+\frac{1}{6}i_XF-\frac{1}{12}\tilde{X}\wedge F.
\end{equation}
Then, the Dirac operator is defined as
\begin{equation}
\displaystyle{\not}{\cal{D}}=\displaystyle{\not}D+\frac{1}{12}F
\end{equation}
and the supergravity twistor spinors are solutions of the following
equation
\begin{equation}
{\cal{D}}_X\psi=\frac{1}{11}\tilde{X}\displaystyle{\not}{\cal{D}}\psi.
\end{equation}

We can calculate the covariant derivative of $p$-form Dirac currents
of supergravity twistor spinors through the above equalities;
\begin{eqnarray}
\nabla_{X_b}\wp_p(a_{\psi})&=&\frac{1}{11}\wp_p(e_b\displaystyle{\not}{\cal{D}}\psi\bar{\psi}+\psi(\overline{e_b\displaystyle{\not}{\cal{D}}\psi}))\nonumber\\
&=&\frac{1}{11}\wp_p\left(e_be^a(\nabla_{X_a}\psi)\bar{\psi}+\frac{1}{12}e_bF\psi\bar{\psi}+\psi\overline{\nabla_{X_a}\psi}(e_be^a)^{\cal{J}}+\frac{1}{12}\psi\overline{F\psi}(e_b)^{\cal{J}}\right)\nonumber\\
&=&\frac{1}{11}\wp_p(e_b\,e^a(\nabla_{X_a}\psi)\bar{\psi}+\psi\overline{\nabla_{X_a}\psi}e^a\,e_b)\nonumber\\
&&+\frac{1}{132}\left[e_b\wedge\wp_{p-1}(F\psi\bar{\psi})+i_{X_b}\wp_{p+1}(F\psi\bar{\psi})\pm e_b\wedge\wp_{p-1}(\psi\overline{F\psi})\pm i_{X_b}\wp_{p+1}(\psi\overline{F\psi})\right]\nonumber\\
&=&\frac{1}{11}e_b\wedge\wp_{p-1}(\displaystyle{\not}da_{\psi}-2e^a\wedge (b_{a})_{\psi})+\frac{1}{11}i_{X_b}\wp_{p+1}(\displaystyle{\not}da_{\psi}-2i_{X^a}(b_{a})_{\psi})\\
&&+\frac{1}{132}\left[e_b\wedge\wp_{p-1}(F\psi\bar{\psi})+i_{X_b}\wp_{p+1}(F\psi\bar{\psi})\pm e_b\wedge\wp_{p-1}(\psi\overline{F\psi})\pm i_{X_b}\wp_{p+1}(\psi\overline{F\psi})\right]\nonumber
\end{eqnarray}
where the terms with $\pm$ signs come from $(e_b)^{\cal{J}}=\pm e_b$ for ${\cal{J}}=\xi$ and ${\cal{J}}=\xi\eta$ respectively. Then, the action of the exterior derivative and co-derivative
operations on $p$-form Dirac currents are obtained as
\begin{eqnarray}
d\wp_p(a_{\psi})&=&\frac{(p+1)}{11}\wp_{p+1}(\displaystyle{\not}da_{\psi}-2i_{X^a}(b_{a})_{\psi})+\frac{(p+1)}{132}\wp_{p+1}(F\psi\bar{\psi}\pm\psi\overline{F\psi})\\
\delta\wp_p(a_{\psi})&=&\frac{(p-12)}{11}\wp_{p-1}(\displaystyle{\not}da_{\psi}-2e^a\wedge (b_{a})_{\psi})+\frac{(p-12)}{132}\wp_{p-1}(F\psi\bar{\psi}\pm\psi\overline{F\psi}).
\end{eqnarray}
The last terms of (58) and (59) are contributions from the
supergravity field $F$. For $F=0$, these equations reduce to (25)
and (26). By comparing (58) and (59) with (57), one can see that the $p$-form Dirac currents of supergravity twistors also satisfy the CKY equation.

On the other hand, supergravity Killing spinors are defined by the following equation
\begin{equation}
{\cal{D}}_X\psi=\lambda\tilde{X}\psi.
\end{equation}
Thus the 4-form field $F$ does not contribute to the covariant
derivative of the $p$-form Dirac currents for supergravity Killing
spinors since the right hand side does not contain $F$ and the covariant derivative of $p$-form Dirac currents becomes same as in Section 3;
\begin{eqnarray}
\nabla_{X_b}\wp_p(a_{\psi})&=&\wp_p\left(({\cal{D}}_{X_b}\psi)\bar{\psi}+\psi{\cal{D}}_{X_b}\bar{\psi}\right)\nonumber\\
&=&\wp_p\left(\lambda e_ba_{\psi}+\lambda^ja_{\psi}(e_b)^{\cal{J}}\right)\nonumber.
\end{eqnarray}
However, the dimension and signature of the spacetime
restricts the spin invariant inner products that can be chosen. In
11 dimensions with 10 positive and 1 negative signs in the metric,
the Clifford bundle of the manifold is constructed from the
complexified Clifford algebra $Cl_{10,1}\otimes_{\mathbb{R}}\mathbb{C}$. By using
the periodicity relations of Clifford algebras \cite{Lawson Michelson}
\begin{eqnarray}
Cl_{r,s+1}&\cong&Cl_{s,r+1}\nonumber\\
Cl_{r,s}&\cong&Cl_{r-4,s+4}
\end{eqnarray}
and from $Cl_{6,5}\cong\mathbb{R}(32)\oplus\mathbb{R}(32)$ and $Cl_{0,1}\cong\mathbb{C}$, we obtain
\begin{eqnarray}
Cl_{10,1}\otimes\mathbb{C}\cong Cl_{6,5}\otimes\mathbb{C}\cong
Cl_{6,5}\otimes
Cl_{0,1}\cong(\mathbb{R}(32)\oplus\mathbb{R}(32))\otimes\mathbb{C}
\end{eqnarray}
where $\mathbb{R}(32)$ denotes the real matrices of $32\times32$
entries. For all Clifford algebras $Cl_{r,s}$ with $r$ positive and
$s$ negative generators, the involutions $\xi$, $\xi\eta$, $\xi^*$
and $\xi\eta^*$ induce one of the ten classes of spin invariant
inner products on it (see Appendix C). From the inner product classification table (see Table 2.15 in \cite{Benn Tucker}),
the involutions $\xi$ and $\xi\eta$ induce the classes shown in Table 4 on
$Cl_{6,5}$ and $Cl_{0,1}$.

\begin{table}[h]
\centering
\begin{tabular}{c|c|c}
  $ $ & $\xi$ & $\xi\eta$ \\ \hline
  $Cl_{6,5}$ & $\mathbb{R}$-swap & $\mathbb{R}$-skew$\oplus\,\mathbb{R}$-skew \\
  $Cl_{0,1}$ & $\mathbb{C}$-sym & $\mathbb{C}^*$-sym \\
  \end{tabular}
\caption{Inner product classes of $Cl_{6,5}$ and $Cl_{0,1}$ induced by $\xi$ and $\xi\eta$.}
\end{table}

The involutions on $Cl_{6,5}\otimes Cl_{0,1}$ are induced by
component algebras by the following rule
\begin{eqnarray}
Cl_{6,5}\otimes Cl_{0,1}&\cong&Cl_{6,5}\otimes\mathbb{C}\nonumber\\
\xi&\cong&\xi\otimes\xi\nonumber\\
\xi\eta&\cong&\xi\eta\otimes\xi\nonumber.
\end{eqnarray}
Hence, the tensor products of inner product classes give \cite{Benn Tucker}
\begin{eqnarray}
\mathbb{R}\text{-swap} \otimes\mathbb{C}\text{-sym}&=&\mathbb{C}\text{-swap} \quad \text{for}
\quad
\xi\nonumber\\
(\mathbb{R}\text{-skew}\oplus\mathbb{R}\text{-skew})\otimes\mathbb{C}\text{-sym}&=&\mathbb{C}\text{-skew}\oplus\mathbb{C}\text{-skew}
\quad \text{for} \quad \xi\eta\nonumber
\end{eqnarray}
for $Cl_{10,1}\otimes \mathbb{C}$. This means that we have the inner
products that correspond to $\{Id,\xi\}$, $\{^*,\xi\}$ and
$\{Id,\xi\eta\}$. The involutions $\xi^*$ and $\xi\eta^*$ induce
$\mathbb{C}$-swap and $\mathbb{C}^*$-sym$\,\oplus\,\mathbb{C}^*$-sym
inner products respectively and these correspond to $\{Id,\xi^*\}$,
$\{^*,\xi^*\}$ and $\{^*,\xi\eta^*\}$. However, they are equivalent to
$\{Id,\xi\}$, $\{^*,\xi\}$ and $\{^*,\xi\eta\}$ on real forms. Hence we
have all four possibilities of inner products that are $\{Id,\xi\}$,
$\{^*,\xi\}$, $\{Id,\xi\eta\}$ and $\{^*,\xi\eta\}$. Thus, the situation
is same as in Section 3 for supergravity Killing spinors. Then, the $p$-form Dirac currents of supergravity Killing spinors also satisfy the KY equation and the properties obtained in Section 3 are also relevant in this case. Especially, 5-form and 6-form Dirac currents of supergravity Killing spinors satisfy the DKP equations;
\begin{eqnarray}
dj^{(5)}_{\psi}=\mu j^{(6)}_{\psi}\quad&,&\quad\delta j^{(5)}_{\psi}=0\nonumber\\
\delta j^{(6)}_{\psi}=-\mu j^{(5)}_{\psi}\quad&,&\quad dj^{(6)}_{\psi}=0\nonumber.
\end{eqnarray}
This corresponds to a coupling of the charges that are induced on 4-branes and 5-branes by Dirac currents.

The action of 11-dimensional supergravity theory is related to the
action of different 10-dimensional supergravity theories one of
which is the type IIA supergravity theory that is the low energy
limit of the type IIA superstring theory. 10-dimensional type IIA
supergravity action can be obtained from the 11-dimensional theory
by Kaluza-Klein reduction. In this method, a Killing vector
$K$ in the 11-dimensional background $(M,g)$ is used for the
dimensional reduction. The new fields dependent on $K$ is obtained
in the 10-dimensional theory and the covariant derivative and Dirac
operators will also dependent on these new fields. Hence, Killing and
twistor spinors are obtained from the equations that are written in
terms of these modified operators.

As an example, we consider the 11-dimensional supergravity
background $(M,g)$ with a Killing vector $K$ and $F=0$. $K$
generates a one-parameter group $\Gamma$ and we have a map
$\pi:M\rightarrow N=M/\Gamma$. The 11-dimensional metric $g$ on $M$
is written in terms of the 10-dimensional metric $h$ on $N$ and the
Killing vector $K$ as follows
\begin{equation}
g=\pi^*h+\tilde{K}\otimes\tilde{K}
\end{equation}
where $\pi^*$ is the pullback map and and $\tilde{K}$ is the 1-form
dual to $K$. The dilaton scalar field $\phi$ and the 2-form field
$F_2$ are defined in terms of $K$ as
\begin{eqnarray}
\exp{(2\pi^*\phi)}&=&g(K,K)\\
\pi^*F_2&=&d\alpha_K
\end{eqnarray}
where $\alpha_K=\tilde{K}/g(K,K)$. Then, $(N,h,F_2,\phi)$ is a
10-dimensional type IIA supergravity background which is the
Kaluza-Klein reduction of $(M,g)$.

The spinor covariant derivative in 10-dimensional supergravity is written as
\begin{equation}
{\cal{D}}_X=\nabla_X-\frac{1}{2}\exp{\phi}\left((i_X\alpha_K)d\phi+\frac{1}{4}i_X F_2\right)z_N
\end{equation}
and the Dirac operator is
\begin{equation}
\displaystyle{\not}{\cal{D}}=\displaystyle{\not}D-\frac{1}{2}\exp{\phi}\left(\alpha_Kd\phi+\frac{1}{2}F_2\right)z_N
\end{equation}
where $z_N$ is the volume form on $N$.

The covariant derivative of $p$-form Dirac currents for supergravity
twistor spinors which are defined in terms of (66) and (67) are
obtained as follows
\begin{eqnarray}
\nabla_{X_b}\wp_p(a_{\psi})&=&\frac{1}{10}\wp_p\left(e_b\displaystyle{\not}{\cal{D}}\psi\bar{\psi}+\psi(\overline{e_b\displaystyle{\not}{\cal{D}}\psi})\right)\nonumber\\
&=&\frac{1}{10}\wp_p\left(e_be^a(\nabla_{X_a}\psi)\bar{\psi}-\frac{1}{2}e_b(\exp\phi)\alpha_K(d\phi)z_N\psi\bar{\psi}-\frac{1}{4}e_b(\exp\phi)F_2z_N\psi\bar{\psi}\right)\nonumber\\
&&+\frac{1}{10}\wp_p\left(\psi\overline{\nabla_{X_a}\psi}(e_be^a)^{\cal{J}}-\frac{1}{2}\psi\overline{(\exp\phi)\alpha_K(d\phi) z_N\psi}(e_b)^{\cal{J}}-\frac{1}{4}\psi\overline{(\exp\phi)F_2z_N\psi}(e_b)^{\cal{J}}\right)\nonumber\\
&=&\frac{1}{10}\wp_p(e_b\,e^a(\nabla_{X_a}\psi)\bar{\psi}+\psi\overline{\nabla_{X_a}\psi}e^a\,e_b)\nonumber\\
&&-\frac{1}{20}\left[\left(e_b\wedge\wp_{p-1}+i_{X_b}\wp_{p+1}\right)\left((\exp{\phi})\alpha_K(d\phi)z_N\psi\bar{\psi}\pm\psi\overline{(\exp{\phi})\alpha_K(d\phi)z_N\psi}\right)\right]\nonumber\\
&&-\frac{1}{40}\left[\left(e_b\wedge\wp_{p-1}+i_{X_b}\wp_{p+1}\right)\left((\exp{\phi})F_2z_N\psi\bar{\psi}\pm\psi\overline{(\exp{\phi})F_2z_N\psi}\right)\right]\nonumber\\
&=&\frac{1}{10}e_b\wedge\wp_{p-1}(\displaystyle{\not}da_{\psi}-2e^a\wedge (b_{a})_{\psi})+\frac{1}{10}i_{X_b}\wp_{p+1}(\displaystyle{\not}da_{\psi}-2i_{X^a}(b_{a})_{\psi})\nonumber\\
&&-\frac{1}{20}\left[\left(e_b\wedge\wp_{p-1}+i_{X_b}\wp_{p+1}\right)\left((\exp{\phi})\alpha_K(d\phi)z_N\psi\bar{\psi}\pm\psi\overline{(\exp{\phi})\alpha_K(d\phi)z_N\psi}\right)\right]\nonumber\\
&&-\frac{1}{40}\left[\left(e_b\wedge\wp_{p-1}+i_{X_b}\wp_{p+1}\right)\left((\exp{\phi})F_2z_N\psi\bar{\psi}\pm\psi\overline{(\exp{\phi})F_2z_N\psi}\right)\right].
\end{eqnarray}
Here $\pm$ terms again correspond to $(e_b)^{\cal{J}}=\pm e_b$ for ${\cal{J}}=\xi$ and ${\cal{J}}=\xi\eta$ respectively. Hence, the action of exterior derivative operator on $p$-form Dirac
currents is
\begin{eqnarray}
d\wp_p(a_{\psi})&=&\frac{(p+1)}{10}\wp_{p+1}(\displaystyle{\not}da_{\psi}-2i_{X^a}(b_{a})_{\psi})\nonumber\\
&-&\frac{(p+1)}{20}\wp_{p+1}\left(e_b(\exp{\phi})\alpha_K(d\phi)z_N\psi\bar{\psi}\pm\psi\overline{(\exp{\phi})\alpha_K(d\phi)z_N\psi}\right)\nonumber\\
&-&\frac{(p+1)}{40}\wp_{p+1}\left(e_b(\exp{\phi})F_2\,z_N\psi\bar{\psi}\pm\psi\overline{(\exp{\phi})F_2z_N\psi}\right)
\end{eqnarray}
and the action of co-derivative operator is
\begin{eqnarray}
\delta\wp_p(a_{\psi})&=&\frac{(p-11)}{10}\wp_{p-1}(\displaystyle{\not}da_{\psi}-2e^a\wedge (b_{a})_{\psi})\nonumber\\
&-&\frac{(p-11)}{20}\wp_{p-1}\left(e_b(\exp{\phi})\alpha_K(d\phi)z_N\psi\bar{\psi}\pm\psi\overline{(\exp{\phi})\alpha_K(d\phi)z_N\psi}\right)\nonumber\\
&-&\frac{(p-11)}{40}\wp_{p-1}\left(e_b(\exp{\phi})F_2\,z_N\psi\bar{\psi}\pm\psi\overline{(\exp{\phi})F_2z_N\psi}\right).
\end{eqnarray}
These differ from the previous identities obtained for ordinary and
11-D supergravity cases in (25-26) and (58-59). However, for $\phi=0$ and $F_2=0$ these identities reduce to (26-27). By comparing (69) and (70) with (68) one can see that $p$-form Dirac currents of 10-dimensional supergravity twistor spinors also satisfy the CKY equation.

For supergravity Killing spinors, we have to consider the choices of spinor inner products again. In 10 dimensions with 9 positive and 1 negative signs in the metric,
the Clifford bundle of the manifold is constructed from the
complexified Clifford algebra $Cl_{9,1}\otimes_{\mathbb{R}}\mathbb{C}$. By using
the periodicity relations of Clifford algebras and from $Cl_{5,5}\cong\mathbb{R}(32)$ we obtain
\begin{eqnarray}
Cl_{9,1}\otimes\mathbb{C}\cong Cl_{5,5}\otimes\mathbb{C}\cong
Cl_{5,5}\otimes Cl_{0,1}\cong\mathbb{R}(32)\otimes\mathbb{C}
\end{eqnarray}
From the inner product classification table \cite{Benn Tucker}, the involutions $\xi$
and $\xi\eta$ induce the classes in Table 5 on $Cl_{5,5}$ and
$Cl_{0,1}$

\begin{table}[h]
\centering
\begin{tabular}{c|c|c}
  $ $ & $\xi$ & $\xi\eta$ \\
  \hline
  $Cl_{5,5}$ & $\mathbb{R}$-sym & $\mathbb{R}$-skew \\
  $Cl_{0,1}$ & $\mathbb{C}$-sym & $\mathbb{C}^*$-sym \\
  \end{tabular}
  \caption{Inner product classes for $Cl_{5,5}$ and $Cl_{0,1}$ induced by $\xi$ and $\xi\eta$.}
\end{table}

The tensor products of inner product classes give \cite{Benn Tucker}
\begin{eqnarray}
\mathbb{R}\text{-sym} \otimes\mathbb{C}\text{-sym}&=&\mathbb{C}\text{-sym} \quad \text{for}
\quad
\xi\nonumber\\
\mathbb{R}\text{-skew}\otimes\mathbb{C}\text{-sym}&=&\mathbb{C}\text{-skew} \quad \text{for}
\quad \xi\eta\nonumber
\end{eqnarray}
in $Cl_{9,1}\otimes\mathbb{C}$. This means that we have the inner
products that correspond to $\{Id,\xi\}$ and $\{Id,\xi\eta\}$. The
involutions $\xi^*$ and $\xi\eta^*$ induce $\mathbb{C}^*$-sym inner
product and this corresponds to $\{^*,\xi^*\}$ and
$\{^*,\xi\eta^*\}$. However, they are equal to $\{^*,\xi\}$ and
$\{^*,\xi\eta\}$ on real forms. Then we have all four possibilities
of inner products that are $\{Id,\xi\}$, $\{^*,\xi\}$,
$\{Id,\xi\eta\}$ and $\{^*,\xi\eta\}$. So, exactly the same
situations are relevant for $p$-form Dirac currents of supergravity
Killing spinors as in 11-dimensional case. Therefore, they satisfy
the KY equation and properties of them are similar to those found in
Section 3. However, they do not satisfy the DKP equations, because
the dimension is even.

\section{Conclusion}

Dirac current of a spinor is defined by a projection operator which
takes a spinor and gives a 1-form. We generalize the definition of
the Dirac current to higher degree forms and define the $p$-form
Dirac currents by using higher degree projection operators that take
a spinor and give a $p$-form. We consider the special cases of
twistor spinors and Killing spinors which are essential ingredients
of supergravity theories in 10 and 11-dimensions. The existence and number of Killing spinors are essential to generate the supersymmetry
transformations in these theories. We found that $p$-form Dirac
currents of twistor spinors correspond to the CKY forms and $p$-form Dirac currents of Killing spinors correspond to the KY
forms of the background spacetime. This means that $p$-form Dirac
currents of twistor spinors and Killing spinors can give way to
define generalized bosonic supercharges in supergravity theories.

In the case of Killing spinors, exterior and co-derivatives of
$p$-form Dirac currents are dependent on the choice of spinor inner
products. There are two main classes of Dirac currents which
correspond to closed co-exact ones and co-closed exact ones. The
dual relation between these two classes gives rise to a resemblance
of a duality in Maxwell equations. In one case, the $p$-form Dirac
currents can be seen as source terms in field equations and in the
other case the Hodge duals of $p$-form Dirac currents are source
terms. This implies a duality between Noether-like charges and
topological charges constructed from $p$-form Dirac currents. On the
other hand, these closedness and co-closedness properties also have
a relation with DKP equations. In some odd dimensions, $p$-form and
$(p-1)$-form Dirac currents correspond to a couple that satisfy DKP
equations and this means that bosonic supercharges on $(p-1)$ branes
and $(p-2)$ branes are coupled through these equations. Moreover,
the $p$-form Dirac currents of Killing spinors are eigenforms of the
Laplace-Beltrami operator and in the parallel spinor  and pure
spinor cases, $p$-form Dirac currents are harmonic forms.

In 11-dimensional supergravity theory, the existence of the bosonic
4-form field $F$ modifies the covariant derivative and Dirac
operator. Hence, the exterior derivative and co-derivative of
$p$-form Dirac currents for supergravity twistor spinors contain the
contribution of $F$. The situation for
supergravity Killing spinors is dependent on the choice of spinor inner product which is related to the properties of the Clifford algebra in that dimension. 10-dimensional supergravity theory which
corresponds to the Kaluza-Klein reduction of the 11-dimensional one
also contains bosonic fields that are the dilaton field $\phi$ and
the 2-form field $F_2$. Existence of these fields modifies the
exterior derivative and co-derivative of $p$-form Dirac currents for
supergravity twistor spinors. The situation for supergravity Killing spinors is same as before. However, in both 10 and 11-dimensional cases the $p$-form Dirac currents of supergravity twistor and Killing spinors correspond to CKY and KY forms of the background respectively.

CKY forms and KY forms generate the symmetry operators for the
massless and massive Dirac equation respectively. Since the
$p$-form Dirac currents of twistor spinors and Killing spinors
correspond to CKY forms and KY forms, they can also be used to
construct symmetry operators for the Dirac equation. Hence, the
existence of twistor and Killing spinors can give rise to find new
Dirac spinors through symmetry operators. On the other hand, KY
forms are used in the construction of gravitational currents \cite{Acik Ertem Onder Vercin2,Ertem Acik}.
This means that the $p$-form Dirac currents of Killing spinors can be related to the
conserved gravitational currents of the underlying background. The
relation between the conserved gravitational currents and the
generalized bosonic supercharges defined through $p$-form Dirac
currents has a worth to investigate.

\begin{appendix}

\section{The algebra of projectors}
Because of the defining relation of Clifford multiplication,
Clifford algebras are $\mathbb{Z}_{2}$-graded algebras and the odd /
even elements are separated by the main involutary automorphism
$\eta$ \footnote{$\eta$ acts on a homogeneous element $a$ as
$\eta(a):= a^\eta:=(-1)^{deg(a)}a$. In even dimensions the Clifford
algebra is central simple and by Noether-Skolem theorem its all
automorphisms are necessarily inner, then it can easily be seen that
$\eta(a):=zaz^{-1}$ i.e. conjugation of $a$ with respect to the
volume form $z$.}. Since the linear structure of Clifford algebras
coincides with exterior algebras of the same dimension, the $\mathbb{Z}$-gradation of exterior algebras can also be used in Clifford algebra calculations. If the operator that projects an inhomogeneous Clifford algebra element to the $\mathbb{Z}$-homogeneous
subspace of degree $p$ is denoted by $\wp_{p}$, then a general element $\Phi$
of the Clifford algebra can be decomposed into its
$\mathbb{Z}$-homogeneous components as
\begin{eqnarray}
\Phi=\sum_{p=0}^n \wp_{p}(\Phi):=\sum_{p=0}^n \Phi_{p}\nonumber
\end{eqnarray}
in $n$-dimensions. The selection of a $g$-orthonormal coframe field $\{e^a\}$ for
generating the smooth local sections of the Clifford bundle $Cl(M,g)$
has many useful advantages, one of which is that the $p$-homogeneous
parts of inhomogeneous elements can be resolved into their
coefficient expansions with respect to a $g_{p}$-orthonormal
$p$-form basis $\{e^{I(p)}\}$ by
\begin{eqnarray}
\wp_{p}(\Phi)=\wp_{0}(\Phi\,e_{I(p)}^{\xi})e^{I(p)}=(\Phi_{p})_{I(p)}\,e^{I(p)}    \nonumber
\end{eqnarray}
where $\xi$ is the main involutary anti-automorphism of the algebra
\footnote{$\xi$ acts on a decomposable Clifford element
$s=x^{1}x^{2}...\,x^{h}$ of length $h$ as
$s^{\xi}=x^{h}...\,x^{2}x^{1}$ and its action on an exterior
$p$-form $\omega$ is given by $\omega^{\xi}=(-1)^{\lfloor
p/2\rfloor}\omega$ where in every case the order of multiplication
being reversed,$\lfloor\;\rfloor$ denotes the floor function which maps its argument to the largest previous integer.} and
$g_{p}$ is the induced metric on the space of $p$-forms \footnote{For
$\Phi,\Psi \in \Gamma Cl(M,g)$ and assuming $M$ to be orientable then
$g_{p}(\Phi_{p}\,,\Psi_{p})\,z:=\Phi_{p}\wedge \ast\Psi_{p}$
which could also be defined as
$g_{p}(\Phi_{p}\,,\Psi_{p})=\wp_{0}(\Phi_{p}\,\Psi_{p}^{\xi})$,
$z:=\ast 1$ is the orientation fixing volume form: the image
of the unit element under the Hodge map. In $n$-dimensions the basis
$\{e^{I(p)}\}$ is defined as $\{e^{i_{1}}\wedge
e^{i_{2}}\wedge...e^{i_{p}}|\,1\leq i_{1}<i_{2}<...<i_{p}\leq
n\}:=\{e^{\textsc{1}(p)},e^{\textsc{2}(p)},...,e^{C(n,p)(p)}\}$ and
the multi-index $I(p)$ is a well-ordered p-tuple of indices
$(i_{1},i_{2},...,i_{p})$ where $C(n,p)$ is the $p$-combination of $n$. If $J(p)$ is another multi-index
$(j_{1},j_{2},...,j_{p})$ then $I(p)> J(p)$ iff the first index of
$I(p)$ different from the corresponding index in $J(p)$ is greater,
i.e., $i_{1}=j_{1},i_{2}=j_{2},...,i_{k-1}=j_{k-1},i_{k}\neq j_{k}$
necessitates $i_{k} > j_{k}$. So the multi-index set
$\mathcal{I}(p):=\{I(p)\}=\{\textsc{1}(p),\textsc{2}(p),...,C(n,p)(p)\}$
is a well-ordered set for $\forall p$ and its cardinality is equal
to the dimension of the $p$-form space that is $C(n,p)$.}. The other main
advantage of choosing orthonormal frames is that the Clifford forms
and exterior forms coincide in this case and the only difference
between the Clifford algebra and the exterior algebra is the algebra
product used in defining them. If $\textsf{g}$ is the coefficient
matrix of $g$ in some chart, the sign of its determinant is defined as
$\varepsilon(\textsf{g}):=\frac{det\textsf{g}}{|det\textsf{g}|}$. Let $\Phi$ be a general section of the Clifford bundle, i.e.,
$\Phi\in \Gamma Cl(M,g)$, $A$ a 1-form $A \in \Gamma T^\ast M$ and the vector field that is metric dual of $A$ is $\widetilde{A}=g(A,.)\in \Gamma TM$ and $\Gamma Cl(M,g)$, $\Gamma T^\ast M$ and $\Gamma TM$ denote the sections of the Clifford bundle, cotangent bundle and tangent bundle respectively, then the following identities hold
\begin{eqnarray}
(\Phi\wedge A)={\ast}^{-1}i_{\widetilde{A}}\ast\Phi=(-1)^{(n-1)}\ast i_{\widetilde{A}}{\ast}^{-1}\Phi
\end{eqnarray}
or
\begin{eqnarray}
(A\wedge \Phi)={\ast}^{-1}i_{\widetilde{A}}\ast\eta\,\Phi=(-1)^{(n-1)}\ast i_{\widetilde{A}}{\ast}^{-1}\eta\,\Phi  \nonumber
\end{eqnarray}
which can be used to rewrite Clifford product as
\begin{eqnarray}
A \Phi=(i_{\widetilde{A}}+{i_{\widetilde{A}}}^\dag)\,\Phi \nonumber
\end{eqnarray}
where ${i_{\widetilde{A}}}^\dag:= {\ast}^{-1}i_{\widetilde{A}}\ast\eta$. The projector operators satisfy the identities listed below;

\begin{eqnarray}
1=\sum_{p=0}^n \wp_{p}
\end{eqnarray}
\begin{eqnarray}
\wp_{p}\,\wp_{q}=\delta_{pq} \,\wp_{p}\qquad(\textrm{no\,sum})
\end{eqnarray}
\begin{eqnarray}
\ast\wp_{p}{\ast}^{-1}=\wp_{n-p}={\ast}^{-1}\wp_{p}\ast
\end{eqnarray}
\begin{eqnarray}
i_{X}\,\wp_{p}=\wp_{p-1}\,i_{X}
\end{eqnarray}
\begin{eqnarray}
\ast\ast\wp_{p}=(-1)^{p(n-p)}\varepsilon(\textsf{g})\wp_{p}=\ast^{-1}\ast^{-1}\wp_{p}.
\end{eqnarray}

Since the projectors commute with degree preserving operations such as
Lie differentiation $\mathfrak{L}_{X}$, covariant differentiation
$\nabla_{X}$ or the maps $\xi$ and $\eta$, the following relations can also be derived;
\begin{eqnarray}
i_{\widetilde{A}}\,\Phi=\ast(({\ast}^{-1}\Phi)\wedge A)=(-1)^{(n-1)}{\ast}^{-1}((\ast\Phi)\wedge A)
\end{eqnarray}
\begin{eqnarray}
A\wedge\wp_{p}=\wp_{p+1}\,A\wedge
\end{eqnarray}
\begin{eqnarray}
\ast i_{\tilde{A}}\wp_{p}=(-1)^{(p+1)}A\wedge\ast\wp_{p}
\end{eqnarray}
\begin{eqnarray}
d\,\wp_{p}=\wp_{p+1}\,d
\end{eqnarray}
\begin{eqnarray}
\delta\,\wp_{p}=\wp_{p-1}\,\delta
\end{eqnarray}
where $d$ is the exterior derivative and $\delta:={\ast}^{-1}d\ast\eta$ is the
coderivative which in the Riemannian case can be defined as the
adjoint of the exterior derivative $d$ with respect to the inner
product $\langle\alpha,\beta\rangle:=\int_{M}\alpha\wedge\ast\beta$
for arbitrary forms $\alpha$ and $\beta$ with the same degree. The first (second) equality in (A7)
can be derived from the first (second) equality of (A1) by the map
$\Phi \mapsto {\ast}^{-1}\Phi\:(\Phi \mapsto \ast\Phi)$ or
vice-versa. Application of (A4)((A5)) and (A5)((A4)) respectively
give (A8)((A9)) with $\widetilde{X}=A$ and using (A8)((A5)) with the
pseudo-Riemannian equality $d=e^{a}\wedge
\nabla_{X_{a}}\:(\delta=-i_{\widetilde{e^{a}}}\nabla_{X_{a}})$ gives
(A10)((A11)). Here $\{X_{a}\}$ is the $g$-orthonormal frame dual to
$\{e^{a}\}$;$\,e^{a}(X_{b})={\delta^{a}}_{b}$. To give an example
for a simple checking, we can use the second equality of (A1),
(A4), (A5), the properties of the wedge product and the automorphism
$\eta$ for obtaining (A8) step by step as follows
\begin{eqnarray}
\wp_{p}(\widetilde{X}\wedge\Phi)&=&\wp_{p}({\Phi}^{\eta}\wedge\widetilde{X})\nonumber\\
&=&(-1)^{(n-1)}\wp_{p}(\ast i_{X}{\ast}^{-1}{\Phi}^{\eta})\nonumber \\
&=&(-1)^{(n-1)}\ast i_{X}\wp_{n-p+1}({\ast}^{-1}{\Phi}^{\eta})\nonumber\\
&=&(-1)^{(n-1)}\ast i_{X}{\ast}^{-1}\wp_{p-1}({\Phi}^{\eta})\nonumber\\
&=&(-1)^{2(n-1)}\ast{\ast}^{-1}[\wp_{p-1}({\Phi}^{\eta})\wedge \widetilde{X}]\nonumber\\
&=&[\wp_{p-1}(\Phi)]^{\eta}\wedge \widetilde{X}\nonumber\\
&=&\widetilde{X}\wedge \wp_{p-1}(\Phi)  \nonumber.
\end{eqnarray}
When we assume
$\Phi$ to be a homogeneous element it can be written as
$\Phi=\Phi_{p}=\wp_{p}\Phi$ for some $0\leq p\leq n$. So, we can particularly use the algebraic interactions of $\wp$ with other operators and deduce the known relations from different ways. For example (A9) can be obtained by (A7)
in this way. An important identity of Clifford calculus not to be passed
without mentioning is
\begin{eqnarray}
\ast \Phi= \Phi^{\xi}\,z \nonumber
\end{eqnarray}
also known as the Clifford-K\"{a}hler-Hodge duality.

\section{Clifford Idempotents and Spin Structures}

Spinors (or semi-spinors) can be defined as objects carrying the irreducible
representations of a Clifford algebra (or Clifford group) and any
such irreducible representation is equivalent to that carried by a
minimal left ideal (MLI) of the Clifford algebra. So, any MLI of
$Cl(V,g)$ can naturally be termed as the space of spinors. However,
for a pseudo-Riemanian manifold $M$ the Clifford bundle $Cl(M,g)$ has
the Clifford algebra of the cotangent space at the
point $m$, namely $Cl(T_{m}^\ast M,g)$ as the fibre, so any MLI of this fibre
algebra carries the spinor representation. If this MLI assignment
for every point of $M$ could smoothly be done, then one would have a
sub-bundle of the Clifford bundle where each fibre carries an
irreducible representation of the corresponding fibre algebra of the
Clifford bundle. But, the requirement of spinor bundle to be a sub-bundle of
the Clifford bundle puts extra constraints on $M$ via global conditions on the primitive Clifford
idempotent fields over $M$. Therefore, rather then taking MLI's as
spinor spaces we only require them to carry a representation
equivalent to that carried by any MLI. This freedom of choice is
based on a celebrated theorem about associative algebras that all irreducible
representations of a simple algebra are equivalent; in the
semi-simple case the irreducible representations are equivalent if
and only if their kernels are the same. If a spin structure exists
on $M$, then any bundle of spinor spaces will locally be isomorphic
to a bundle of MLI's which is a sub-bundle of the Clifford bundle.
This local opportunity identifies spinor fields with differential
forms that are sections of the local MLI bundle and also this
identification permits us the full usage of the algebraic power for
local calculations.

A \textit{MLI spinor-structure} is obtained when both a usual spinor
structure and an algebraic spinor structure exists simultaneously on $M$. A
\textit{usual spinor structure} (US) puts a global topological
constraint known as the vanishing of the second Stiefel-Whitney
class of the tangent bundle $TM$: $w_{2}(TM)=0$, so a well
defined spinor connection can be obtained for parallel transporting
spinor fields on $M$ and physically the particle spectrum of the
spacetime manifold $M$ is controlled by its topology to
allow fermions.

An \textit{algebraic spin structure} (AS) is a globally defined
primitive idempotent $P$ generating a MLI bundle which is a
sub-bundle of the Clifford bundle. An algebraic spinor is a
(generally inhomogeneous) element $\Phi$ of the exterior bundle
subject to the algebraic constraint
\begin{eqnarray}
\Phi P =\Phi  \nonumber
\end{eqnarray}
which means that $\Phi$ is a smooth section of the minimal left
ideal generated by the primitive $P$, i.e. $\Phi \in \Gamma
I_{L}(P)$. The associativity of Clifford product guaranties that
every left Clifford multiple of an algebraic spinor is also an
algebraic spinor, this is the restatement of saying that left
Clifford ideals are left Clifford modules. The above equation
annihilates the redundant degrees of freedom (dof) for the field
$\Phi$  subject to physical field equations. If this dof reduced
$\Phi$ also solves the neutral K\"{a}hler equation
\begin{eqnarray}
\displaystyle{\not}d \Phi =m \Phi  \nonumber
\end{eqnarray}
with the condition $\nabla_{X}P=0, \forall X \in \Gamma TM $ then in
this special case this \textit{K\"{a}hler spinor} is termed as an \textit{algebraic Dirac
spinor} where $\displaystyle{\not}d=d-\delta$ is the Hodge-de Rham operator. The condition of 'being parallel' on the global primitive $P$ is
necessary for the connection $\nabla$ to preserve the MLI projected
by $P$ which also restricts the geometry of $M$ \cite{Al Saad Benn}.

\section{Spinor Inner Products}
A spin-invariant inner product can be constructed for spinors $\phi$ and $\psi$ as follows
\begin{equation}
\left(\phi,\psi\right)=J^{-1}\phi^{{\cal{J}}}\psi
\end{equation}
where ${\cal{J}}$ is an involution of the algebra that can be $\xi$, $\xi\eta$, $\xi^*$ or $\xi\eta^*$, $J$ is an arbitrary element of the algebra that has the property $P^{{\cal{J}}}=JPJ^{-1}$ and $J^{\cal{J}}=\epsilon J$ with $\epsilon=\pm1$ and $P$ is the primitive idempotent. A Clifford algebra can be simple or semi-simple, namely its structure is in the form of $\mathbb{D}(n)$ or $\mathbb{D}(n)\oplus\mathbb{D}(n)$ where $\mathbb{D}$ is a real division algebra such that $\mathbb{D}\cong\mathbb{R},\mathbb{C}$ or $\mathbb{H}$ and $\mathbb{D}(n)$ is the $n\times n$ matrix algebra whose entries are in $\mathbb{D}$. In simple cases, the result of the inner product will be an element of $\mathbb{D}$ and for any Clifford algebra element $a$ and $q\in\mathbb{D}$ it has the following properties
\begin{eqnarray}
\left(\phi,a\psi\right)&=&\left(a^{\cal{J}}\phi,\psi\right)\\
\left(\phi,\psi q\right)&=&\left(\phi,\psi\right)q\\
\left(\phi q,\psi\right)&=&q^j\left(\phi,\psi\right)
\end{eqnarray}
where $q^j=J^{-1}q^{{\cal{J}}}J$, hence $j$ is an involution of $\mathbb{D}$ and it can be equivalent to identity (Id) in $\mathbb{D}\cong\mathbb{R}$, identity or complex conjugation ($^*$) in $\mathbb{D}\cong\mathbb{C}$, quaternionic conjugation ($\,{\bar{   }}\,$) or reversion ($\,{\hat{   }}\,$) in $\mathbb{D}\cong\mathbb{H}$. Under the exchange of $\phi$ and $\psi$, the inner product can be written as
\begin{equation}
\left(\psi,\phi\right)=\epsilon\left(\phi,\psi\right)^j.
\end{equation}
Such a product will be called $\mathbb{D}^j$-symmetric or $\mathbb{D}^j$-skew for $\epsilon$ is +1 or -1. In semi-simple cases, the result of the inner product will be an element of $\mathbb{D}\oplus\mathbb{D}$ and for any Clifford algebra element $a$ and for $q\in\mathbb{D}\oplus\mathbb{D}$ it has the properties
\begin{eqnarray}
\left(\phi,a\psi\right)&=&\left(a^{\cal{J}}\phi,\psi\right)\\
\left(\phi q,\psi\right)&=&q^{{\cal{J}}}\left(\phi,\psi\right)\\
\left(\psi,\phi\right)&=&\left(\phi,\psi\right)^{{\cal{J}}}.
\end{eqnarray}
For an involution ${\cal{J}}$ that does not preserve the simple components, the inner product will be called $\mathbb{D}$-swap, otherwise it will be $\mathbb{D}$-sym$\,\oplus\,\mathbb{D}$-sym or $\mathbb{D}$-skew$\oplus\,\mathbb{D}$-skew.

From the above considerations, one can obtain that there are ten different spin invariant inner product classes which are listed in Table 6 \cite{Benn Tucker}.

\begin{table}[h]
\centering
\begin{tabular}{c|c c c|c}

  $1$ & $\mathbb{R}\text{\,-sym}$& &$6$&$\mathbb{H}\,{\bar{   }}\text{\,-sym}$ \\
  $2$ & $\mathbb{R}\text{\,-skew}$& &$7$ & $\mathbb{H}\,{\hat{   }}\text{\,-sym}$ \\
  $3$ & $\mathbb{C}\text{\,-sym}$ & &$8$ & $\mathbb{R}\text{\,-swap}$ \\
  $4$ & $\mathbb{C}\text{\,-skew}$& &$9$ & $\mathbb{H}\text{\,-swap}$ \\
  $5$ & $\mathbb{C}^{\,*}\text{\,-sym}$& &$10$ & $\mathbb{C}\text{\,-swap}$ \\

  \end{tabular}
  \caption{Classes of spin invariant inner products.}
\end{table}

\end{appendix}

\begin{acknowledgments}
This work was supported in part by the Scientific and Technological
Research Council of Turkey (T\"{U}B\.{I}TAK) grant 113F103.
\end{acknowledgments}


\end{document}